\documentstyle[12pt,epsf,amssymbols]{article}
\newcommand{\preprint}[1]{\begin{table}[t]  
            \begin{flushright}              
            \begin{large}{#1}               
            \end{large}    
            \end{flushright}                
            \end{table}}                           
\preprint{TAUP-2420-97}
\tiny\normalsize
\begin{document}
\title{Bi-Graded Markovian Matrices as Non-Local Dirac Operators and a New Quantum Evolution.}
\author{E. Atzmon\thanks{%
atzmon@post.tau.ac.il}  \\ 
Raymond and Beverly Sackler Faculty of Exact Sciences,\\ School of Physics
and Astronomy.\\ Tel - Aviv University. }
\date{December 21, 1996}
\maketitle

\begin{abstract}
Measuring distances on a lattice in noncommutative geometry involves
square, symmetric and real ``three-diagonal'' matrices, with the sum
of their elements obeying a supremum condition, together with a
constraint forcing the absolute value of the maximal eigenvalue to be
equal to $1$. In even dimensions, these matrices are unipotent of order
two, while in odd dimensions only their squares are Markovian. We suggest
that these bi-graded Markovian matrices (i.e. consisting in the square
roots of Markovian matrices) can be thought of as non-local Dirac
operators. The eigenvectors of these matrices are spinors. Treating
these matrices as determining the stochastic time evolution of states
might explain why one observes only left handed neutrinos. Some other
physical interpretations are suggested. We end by presenting a
mathematical conjecture applying to $q-\;$graded Markovian matrices.
\end{abstract}

\newpage

\section{ Bi-graded Markovian matrices as non-local Dirac operators }
Measuring distances in noncommutative geometry involves the spectral
triple $(A,\;H,\;D)$. Where $A-\; $is the algebra of functions
defined over the space, $H-\; $ is a Hilbert space of spinors, and
$D-\; $ is a Dirac operator. The Dirac operator acts essentially on
both the functional space and the Hilbert space. The distance in
noncommutative geometry is defined by the following formula:
\begin{equation}
\label{eq.2}d(a,b)=\sup _f\left\{ \left| f(a)-f(b)\right| \;:\;f\in
A,\;\left\| \;\left[ \;D,f\;\right] \;\right\| \leq 1\right\}
\end{equation} 
This formula was applied to an infinite one-dimensional lattice. The
lattice is defined as:\,
$\left\{x_k\;=\;ka\;,\;\;\;k\;\in {\Bbb Z}\right\}$, where $a$ is
the lattice constant (i.e. the quantity which carries the relevant
physical unit). Finding the distances on a one-dimensional lattice was
the goal of \cite{1,2}. In these works \cite{1,2}, one uses the local
discrete Dirac-Wilson operator, usually applied in lattice gauge
theories\footnote{In \cite{1,2} the Dirac-Wilson operator used was the
discrete differentiation, defined as $\frac {f_{i+1}-f_{i-1}}{2a}$}.
The $\left\| \;\left[ \;D,f\;\right] \;\right\| \leq 1$ condition and the
$\sup_f\left| f(a)-f(b)\right|$ condition of (eq.\ref{eq.2}) are here 
reformulated for the evaluation of real, square and symmetric
``three-diagonal'' matrices $M_k$ of the following form:\\
\begin{equation}
\label{eq.8}M_k = \left(
\begin{array}{ccccc}
0 & \Delta _1 & 0 & 0 & 0 \\
\Delta _1 & 0 & \Delta _2 & 0 & 0 \\
0 & \Delta _2 & 0 & \ddots & 0 \\
0 & 0 & \ddots & \ddots & \Delta _{k-1} \\
0 & 0 & 0 & \Delta _{k-1} & 0
\end{array}
\right)
\end{equation}
under the restrictions that the $\sum_i\triangle _i$ should be a supremum
and that the maximal eigenvalue of $M_k - \;\;\; \lambda_{max} \le 1$,
where $\Delta _{i}=f_i\;-\;f_{i-1}$, with the norm of a matrix defined as
the largest eigenvalue. The distance from some $0-\;$site to the
$(k-1)-\;$site of the lattice is $2a\sum_i\triangle _i $.\\
In this work, we assign to the matrix $M_k$ as a whole an interpretation
as a non-local Dirac operator. This interpretation will be further
justified in the sequel, but we note, meanwhile, that $M_k$ is constructed
solely out of local Dirac-Wilson operators, acting both on the functional
space and on the Hilbert space defined over the lattice. As the matrix
$M_k$ acts simultaneously on different sites (essentially $k$ sites), 
it should be considered as a non-local operator.

The problem of finding the distances on a lattice was formulated in
\cite{1} and resolved in \cite{1,2}. What was proved in \cite{2} is that
the matrices are given (under the above restrictions) by the following
formulae:\\

$k=2n   :\,\,\,\,\,\,\triangle _{2i-1}^{\left( k\right) } = 1\,\,\,,\,\,\,\triangle _{2i}^{\left( k\right) } = 0 $\\

$ k=2n+1 :\,\,\,\,\,\,$            
\begin{equation}
\label{eq.20}\triangle _i^{\left( k\right) }=\frac{\frac 12\left( 1-\left(
-1\right) ^i\right) \left( k+1\right) + 2\left( -1\right) ^i\left[
\frac{i+1}2\right] }{\sqrt{\left(k-1\right)\left( k+1\right) }}%
\;\;\;\;\forall \;i\leq k-1
\end{equation}
where the\ $\left[ \;\right] \;\;$brackets stand for the integer value of
the term within.\\
The numbers for the first few odd cases are of the form:\\

$k=3\;:\;\;\;\; \left\{ \frac {1}{\sqrt{2}} , \frac {1}{\sqrt{2}} \right\}$\\

$k=5\;:\;\;\;\; \left\{ \frac {2}{\sqrt{6}} , \frac {1}{\sqrt{6}} , \frac {1}{\sqrt{6}} , \frac {2}{\sqrt{6}} \right\} $\\

$k=7\;:\;\;\;\; \left\{ \frac {3}{\sqrt{12}} , \frac {1}{\sqrt{12}} , \frac {2}{\sqrt{12}} , \frac {2}{\sqrt{12}} , \frac {1}{\sqrt{12}} , \frac {3}{\sqrt{12}}   \right\}$\\

$k=9\;:\;\;\;\; \left\{ \frac {4}{\sqrt{20}} , \frac {1}{\sqrt{20}} , \frac {3}{\sqrt{20}} , \frac {2}{\sqrt{20}} , \frac {2}{\sqrt{20}} , \frac {3}{\sqrt{20}} , \frac {1}{\sqrt{20}} , \frac {4}{\sqrt{20}}   \right\}$\\

As one can see, those sequences alternate in a very special way. This is
illustrated for $k=41$ in the following figure, (with a slight smoothening
out):

\vspace{10mm}
\epsfxsize=11truecm
\centerline{\epsfbox{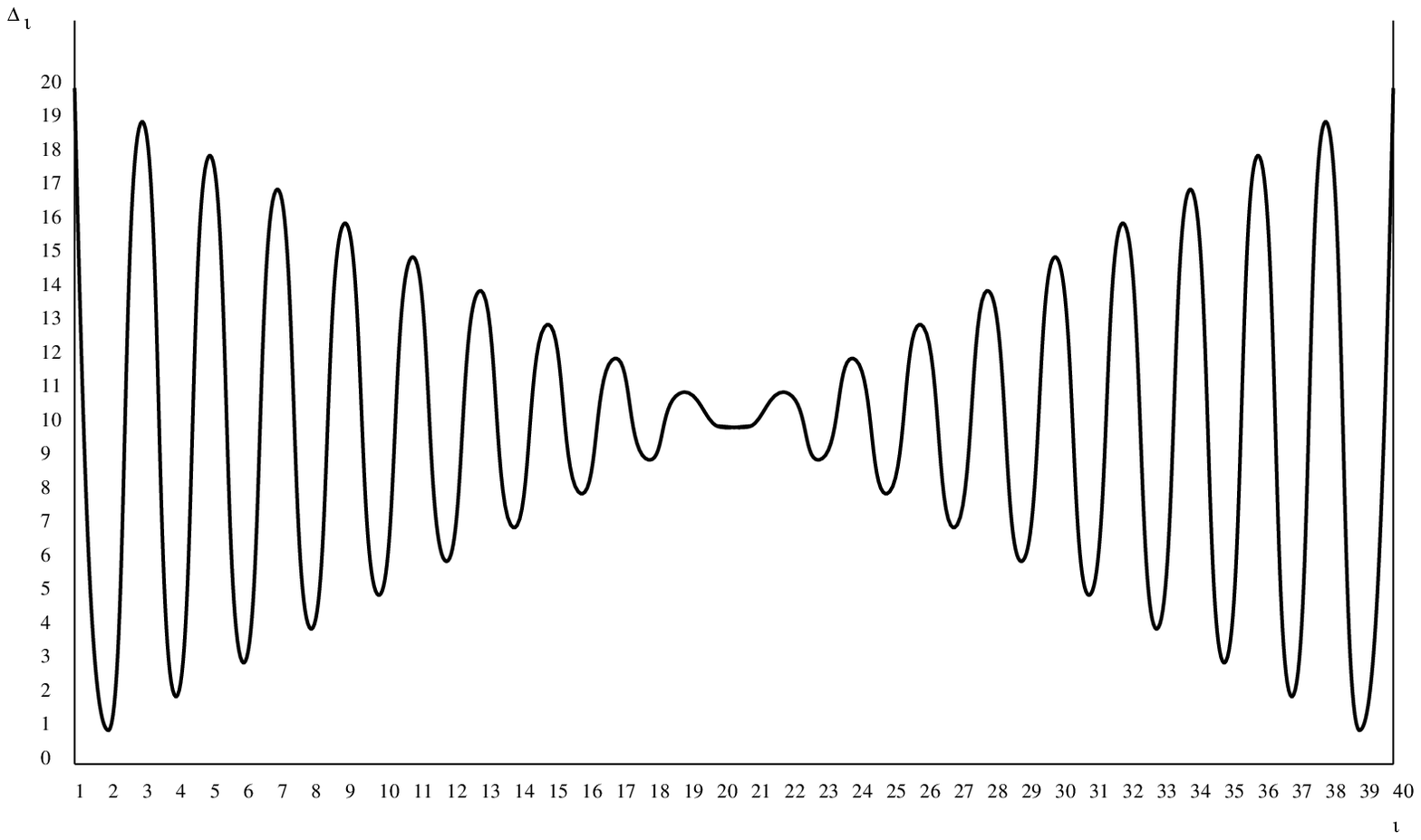}}

\vspace{2mm}
\centerline{\parbox{11truecm}{Figure 1.{\footnotesize The values of $\Delta _i$ for $k=41$ in units of $\frac 1{\sqrt {420}}$.}}}
\vspace{5mm}

These matrices have maximal eigenvalue 1, and the $\sum_i\triangle _i$ is a
supremum. For $k $ even,  the matrices are unipotent of order two
(therefore also Markovian). For $k $ odd, however, only their squares are
Markovian. We demonstrate these features for the the first few matrices: \\
$k=2 :$\\
\[
\left( 
\begin{array}{cc}
0 & 1 \\ 
1 & 0 \\
\end{array}
\right) ^{2n}=\left( 
\begin{array}{cc}
1 & 0 \\ 
0 & 1 \\
\end{array}
\right) 
\]\\
\[
\left( 
\begin{array}{cc}
0 & 1 \\ 
1 & 0 \\
\end{array}
\right) ^{2n+1}=\left( 
\begin{array}{cc}
0 & 1 \\ 
1 & 0 \\
\end{array}
\right) 
\]\\
$k=3 :$
\[
\left( 
\begin{array}{ccc}
0 & \frac 1{\sqrt{2}} & 0 \\ 
\frac 1{\sqrt{2}} & 0 & \frac 1{\sqrt{2}} \\ 
0 & \frac 1{\sqrt{2}} & 0
\end{array}
\right) ^ {2n} =
\left( 
\begin{array}{ccc}
\frac 12 & 0 & \frac 12 \\ 
0 & 1 & 0 \\ 
\frac 12 & 0 & \frac 12
\end{array}
\right) \] 
\newline 
\[
\left( 
\begin{array}{ccc}
0 & \frac 1{\sqrt{2}} & 0 \\ 
\frac 1{\sqrt{2}} & 0 & \frac 1{\sqrt{2}} \\ 
0 & \frac 1{\sqrt{2}} & 0
\end{array}
\right) ^{2n+1}=\left( 
\begin{array}{ccc}
0 & \frac 1{\sqrt 2} & 0 \\ 
\frac 1{\sqrt 2} & 0 & \frac 1{\sqrt 2} \\ 
0 & \frac 1{\sqrt 2} & 0 \\
\end{array}
\right) 
\]\\
$k=4 :$\\
\[
\left( 
\begin{array}{cccc}
0 & 1 & 0 & 0 \\ 
1 & 0 & 0 & 0 \\
0 & 0 & 0 & 1 \\
0 & 0 & 1 & 0 \\
\end{array}
\right) ^{2n}=\left( 
\begin{array}{cccc}
1 & 0 & 0 & 0 \\ 
0 & 1 & 0 & 0 \\
0 & 0 & 1 & 0 \\
0 & 0 & 0 & 1 \\
\end{array}
\right) 
\]\\
\[
\left( 
\begin{array}{cccc}
0 & 1 & 0 & 0 \\ 
1 & 0 & 0 & 0 \\
0 & 0 & 0 & 1 \\
0 & 0 & 1 & 0 \\
\end{array}
\right) ^{2n+1}=\left( 
\begin{array}{cccc}
0 & 1 & 0 & 0 \\ 
1 & 0 & 0 & 0 \\
0 & 0 & 0 & 1 \\
0 & 0 & 1 & 0 \\
\end{array}
\right) 
\]\\

$k=5 :$\\
\[
\left( 
\begin{array}{ccccc}
0 & \frac 2{\sqrt{6}} & 0 & 0 & 0 \\ 
\frac 2{\sqrt{6}} & 0 & \frac 1{\sqrt{6}} & 0 & 0 \\ 
0 & \frac 1{\sqrt{6}} & 0 & \frac 1{\sqrt{6}} & 0 \\
0 & 0 & \frac 1{\sqrt{6}} & 0 & \frac 2{\sqrt{6}} \\
0 & 0 & 0 & \frac 2{\sqrt{6}} & 0 \\
\end{array}
\right) ^2 =
\left( 
\begin{array}{ccccc}
\frac 23 & 0 & \frac 13 & 0 & 0 \\ 
0 & \frac 56 & 0 & \frac 16 & 0 \\
\frac 13 & 0 & \frac 13 & 0 & \frac 13 \\ 
0 & \frac 16 & 0 & \frac 56 & 0 \\
0 & 0 & \frac 13 & 0 & \frac 23 \\
\end{array}
\right) 
\]\\
\[
\left( 
\begin{array}{ccccc}
0 & \frac 2{\sqrt{6}} & 0 & 0 & 0 \\ 
\frac 2{\sqrt{6}} & 0 & \frac 1{\sqrt{6}} & 0 & 0 \\ 
0 & \frac 1{\sqrt{6}} & 0 & \frac 1{\sqrt{6}} & 0 \\
0 & 0 & \frac 1{\sqrt{6}} & 0 & \frac 2{\sqrt{6}} \\
0 & 0 & 0 & \frac 2{\sqrt{6}} & 0 \\
\end{array}
\right) ^{2n \rightarrow \infty }=\left( 
\begin{array}{ccccc}
\frac 13 & 0 & \frac 13 & 0 & \frac 13 \\ 
0 & \frac 12 & 0 & \frac 12 & 0 \\
\frac 13 & 0 & \frac 13 & 0 & \frac 13 \\ 
0 & \frac 12 & 0 & \frac 12 & 0 \\
\frac 13 & 0 & \frac 13 & 0 & \frac 13 \\
\end{array}
\right) 
\]\\
\[
\left( 
\begin{array}{ccccc}
0 & \frac 2{\sqrt{6}} & 0 & 0 & 0 \\ 
\frac 2{\sqrt{6}} & 0 & \frac 1{\sqrt{6}} & 0 & 0 \\ 
0 & \frac 1{\sqrt{6}} & 0 & \frac 1{\sqrt{6}} & 0 \\
0 & 0 & \frac 1{\sqrt{6}} & 0 & \frac 2{\sqrt{6}} \\
0 & 0 & 0 & \frac 2{\sqrt{6}} & 0 \\
\end{array}
\right) ^{2n+1 \rightarrow \infty }=\left( 
\begin{array}{ccccc} 
0 & \frac 1{\sqrt 6} & 0 & \frac 1{\sqrt 6} & 0 \\
\frac 1{\sqrt 6} & 0 & \frac 1{\sqrt 6} & 0 &  \frac 1{\sqrt 6} \\
0 & \frac 1{\sqrt 6} & 0 & \frac 1{\sqrt 6} & 0 \\
\frac 1{\sqrt 6} & 0 & \frac 1{\sqrt 6} & 0 &  \frac 1{\sqrt 6} \\
0 & \frac 1{\sqrt 6} & 0 & \frac 1{\sqrt 6} & 0 \\
\end{array}
\right) 
\]\\
\\
In the general case it will be:\\
\,
$k=even:$\\
\[ 
\begin{array}{c}
\left(M_k\right)^{2n} = {\Bbb I}_k\\
\\
\\
\left(M_k\right)^{2n+1} = M_k\\
\end{array}
\]\\
$k=odd\,\,-$\,These will have an asymptotic behavior:\\
\[
\left(M_k\right)^{2n \rightarrow \infty} =
\left( 
\begin{array}{ccccccc}
\frac {2}{k+1} & 0 & \frac {2}{k+1} & 0 & \frac {2}{k+1} & \ldots & \frac {2}{k+1} \\
0 & \frac 2{k-1} & 0 & \frac 2{k-1} & 0 & \ldots & 0 \\
\frac {2}{k+1} & 0 & \frac {2}{k+1} & 0 & \frac {2}{k+1} & \ldots & \frac {2}{k+1} \\
0 & \frac 2{k-1} & 0 & \frac 2{k-1} & 0 & \ldots & 0 \\
\vdots & \vdots & \vdots & \vdots & \vdots & \vdots  \vdots  \vdots & \vdots \\

0 & \frac 2{k-1} & 0 & \frac 2{k-1} & 0 & \ldots & 0\\
\frac {2}{k+1} & 0 & \frac {2}{k+1} & 0 & \frac {2}{k+1} & \ldots & \frac {2}{k+1} \\
\end{array}
\right)
\]
\newline
\newline
\newline
\[
\left(M_k \right)^{2n+1 \rightarrow \infty} =
\left( 
\begin{array}{cccccc}
0 & \frac 2{\sqrt {(k^2-1)}} & 0 & \frac 2{\sqrt {(k^2-1)}} & \ldots & 0 \\
\frac 2{\sqrt {(k^2-1)}} & 0 & \frac 2{\sqrt {(k^2-1)}} & 0 & \ldots & \frac 2{\sqrt {(k^2-1)}} \\
0 & \frac 2{\sqrt {(k^2-1)}} & 0 & \frac 2{\sqrt {(k^2-1)}} & \ldots & 0 \\
\frac 2{\sqrt {(k^2-1)}} & 0 & \frac 2{\sqrt {(k^2-1)}} & 0 & \ldots & \frac 2{\sqrt {(k^2-1)}} \\
\vdots & \vdots & \vdots & \vdots & \vdots  \vdots  \vdots & \vdots \\
\frac 2{\sqrt {(k^2-1)}} & 0 & \frac 2{\sqrt {(k^2-1)}} & 0 & \ldots & \frac 2{\sqrt {(k^2-1)}} \\
0 & \frac 2{\sqrt {(k^2-1)}} & 0 & \frac 2{\sqrt {(k^2-1)}} & \ldots & 0 \\
\end{array}
\right)
\]\\
\\
The eigenvalues of a bi-graded Markovian matrix are bounded by
$-1\le\lambda_i \le 1$, where the $\pm 1$ eigenvalues always exist.
In the $\left(k=2n\right)-$ case, the $\pm 1$ are the only
eigenvalues which exist (each of them with $n-$ degeneracy). In the
odd case, the $\pm 1$ eigenvalues appear with no degeneracy.\\
The corresponding eigenvectors of the $\pm 1$ eigenvalues have the
following normalized form in the $\left(k=2n\right)-$
case:\footnote{See Appendix B.}
\begin{equation}
\begin{array}{ccccc}
\left\{\pm 1 \leftrightarrow \left(\pm \frac 1{\sqrt{2}}\;,\;\frac
    1{\sqrt{2}}\;,\;0\;,\;\ldots \;,\; 0\right)\right\}\\ 
\vdots \\
\left\{\pm 1 \leftrightarrow \left(0\;,\;\ldots\;,\;0\;,\;\pm \frac 1{\sqrt{2}}\;,\;\frac
    1{\sqrt{2}}\;,\;0\;,\;\ldots \;,\; 0\right)\right\}\\ 
\vdots \\
\left\{\pm 1 \leftrightarrow \left(0\;,\;\ldots\;,\;0\;,\;\pm \frac
    1{\sqrt{2}}\;,\;\frac 1{\sqrt{2}}\right)\right\}
\end{array}
\end{equation}  
The corresponding eigenvectors of the $\pm 1$ eigenvalues have the following
normalized form in the $\left(k=2n+1\right)-$ case:\footnote{See Appendix A.}
\begin{equation}
\begin{array}{lll}
\left\{1 \leftrightarrow \left(\frac {1}{\sqrt{k+1}}\;,\; \frac {1}{\sqrt{k-1}}\;,\; \ldots \; \frac {1}{\sqrt{k-1}}\;,\; \frac
      {1}{\sqrt{k+1}} \right)\right\}\\ 
\\
\left\{-1 \leftrightarrow \left(\frac {\left(-1\right)^n}{\sqrt{k+1}}\;,\; \frac {\left(-1\right)^{n+1}}{\sqrt{k-1}} \;,\; \ldots \;,\; \frac {\left(-1\right)^{n+1}}{\sqrt{k-1}} \;,\; \frac {\left(-1\right)^n}{\sqrt{k+1}} \right)\right\}
\end{array}
\end{equation}\\
Some additional regularities appear for eigenvectors which correspond
to eigenvalues other than $\pm 1$. Example \,\, 1. The sum of all entries 
for each eigenvector is equal to zero. The implication is that all other
eigenvectors are orthogonal to a vector with $1$ in all its entries.
 \,\, 2. The entries in the numerator of the eigenvector corresponding
to the eigenvalue zero are essentially the numbers appearing in a Pascal
triangle, but with alternating signs, separated by zeros.\footnote{See
Appendix A.}
\,3. The eigenvalues of the bi-graded Markovian matrix $M_{2n+1}$
are:\footnote{These eigenvalues are identical with
$\pm\sqrt{\frac{\left<j,m\right.\left|\right.J_-J_+\left.\right|j,m\left.\right>}{\left<j,m\right.\left|\right.J^2\left.\right|j,m\left.\right>}}$
in the case of angular momentum in three dimensions, where
$j\in{\Bbb N},\;\;\;0\le m \le j$.
This should be interpreted as a geometrical average over clockwise (i.e.
holomorphic) and anti-clockwise (i.e. anti-holomorphic) circular motion.}
\\
\begin{equation}
\lambda_{2n+1,\;\pm i}=\pm \sqrt{1 - \frac {i\left(i+1\right)}{n\left(n+1\right)}}\;\;\;\;\;\;\; \forall\;\; 0\;\le\; i \;\le\; n
\end{equation}\\
The eigenvectors, being those of a (nonlocal) Dirac operator, are all
spinors. One should thus not be surprised to observe the eigenvalues all
appearing in pairs which only differ by their sign. This means that for
every spinor, there also appears its anti-spinor (which essentially has
the opposite momentum). In the case of eigenvectors which
correspond to the zero eigenvalue, the spinor and its `anti-spinor' are the
same.

\section{Some possible physical applications of bi-graded Markovian matrices}
A Markovian matrix is essentially a matrix which represents a probabilistic
distribution for transitions per unit ``step'' (possibly, a time interval),
given the initial state. Each row of a Markovian matrix represents some
initial state, while the columns represent the final states. The $(i,j)$
matrix element in a Markovian matrix reproduces the probability for the
system, initially in the $i$-state, to transfer to the $j$-state in one
time interval. The sum of all the probabilities in each row is equal to 1.
Finding the transition probabilities after $n$ time steps is achieved by
exponentiating the Markovian matrix to the $n$-th power. In some cases,
the long-term behavior (i.e. $M^n$ as $n\rightarrow\infty$) becomes 
stationary.\\
In the present case, starting from noncommutative geometry, we found 
bi-graded Markovian matrices. As one can see, in both odd and even cases,
the system oscillates between two phases - which is why we describe the
matrices as bi-graded. However, there is some difference between the odd
and even cases. The even case oscillates between two states, both of which
are Markovian, whereas the odd case oscillates asymptotically
only\footnote{Except in the $n=3$ case, where the oscillations are 
always between two states, of which only one is Markovian.}, and in
such a way that only one of the resulting states is Markovian.\\

Re-examining again the bi-graded matrices $M$, one sees that they can be
decomposed into the sum of two matrices $M_+$ and $M_-$, where $M_+$
includes only the upper off-diagonal part and $M_-$ is its transpose.
The action of $M_{\{+/-\}}$ on a vector will be similar to that of 
raising/lowering operators. It turns out that, in the even case, the
$M_{\{+/-\}}$ operators obey an anti-commutation algebra (\,i.e.
$M_+M_-+M_-M_+={\Bbb I}$\,). However, in the odd case, the
$M_{\{+/-\}}$ obey the commutation relations of a quantum algebra:
\,\,$M_+M_--QM_-M_+=\alpha{\Bbb I}$, \,\,where $Q$ is a rational and
diagonal matrix and $\alpha$ is a rational number, or a Yang-Baxter
like algebra $RM_+M_-=M_-M_+R$.\,\footnote{See Appendix D.}\\

In the following section we suggest an application in which such matrices
are interpreted as describing the Markovian time evolution of the quantum
states of a system.\\

As an example for the even case one considers a one dimensional lattice
with $2n$ lattice sites. We assume that at every odd site there is a spin
$J=1/2$ particle, with $J_{3}=+1/2$, and at every even site there is a
$J=1/2$ particle with $J_{3}=-1/2$ (all the particles are identified).
This state is represented by $\left\{1\;,\;-1\;\ldots\;1\;,\;-1\right\}$. \\
Inspecting $M_{2n}$ and treating it as a Markovian transition matrix
which acts iteratively on states corresponding to the eigenvalue $-1$,
we realize that the lattice is essentially split into pairs of nearest
sites, with the two spin half particles (with opposite orientations)
exchanging  places. An alternative interpretation would consist in
regarding the system as a whole as a description of a stationary spin
wave.\\

Unlike the even case, where $M_{2n}$ itself is Markovian, in the odd case,
$M_{2n+1}$ is not a Markovian matrix; instead, it is the square root of a
Markovian matrix (which is why we have chosen to denote them as bi-graded
Markovian). Thus, the entries in $M_{2n+1}$ do not represent transition
probabilities. However, in $M_{2n+1}^2$ the entries do represent such
transition probabilities. Thus, in analogy to quantum mechanics, one can
identify the $i\;-$ row in the bi-graded Markovian matrix as a {\it ket}
- i.e. $\left.\right|\psi_i\left>\right.$ and the $j\;-$ column as a
{\it bra} - i.e. a $\left<\psi_j\right.\left|\right.$. When the bi-graded
Markovian matrix is squared, the $(i,j)\;-$ entry is
$\left<\psi_j\right.\left.\right|\psi_i\left>\right.$, fitting a
probabilistic interpretation. The bi-graded Markovian matrix as a whole
can thus be thought of as a mixture of quantum states. The main point is
that, unlike conventional quantum mechanics, where the time evolution of
a quantum state is determined by the Hamiltonian, in this interpretation
{\it the bi-graded Markovian matrices represent both the initial state and
its Markovian time evolution}.\\
Heuristically, comparing the role of the Hamiltonian in quantum mechanics
with that of a Markovian matrix in a stochastic process, we are led to the
identification of the square root of a Markovian matrix (i.e. the bi-graded
Markovian matrix) with a (non-local) Dirac operator. It is instructive to
observe the destructive interference and the probability flows in the
Markovian time evolution, in the case of bi-graded Markovian matrices.\footnote{See appendix C.} It turns out that in the asymptotic regime, two
states coexist - i.e. the one which associates zero probabilities to even
sites, and the other, which associates zero probabilities to odd sites.\\

Another possible approach would consist in treating the bi-graded matrices
as operators determining the time evolution of a system, as represented by
a state-vector.\\
Due to the fact that the eigenvectors of the bi-graded matrix span a $k-$
dimensional vector space (the $k-$ dimensional Hilbert space of the spinors)
the state-vector can be represented in that base. When the bi-graded matrix
is exponentiated to some power $m$, its eigenvalues are thereby exponentiated
to the same power $m$. However, all the eigenvalues satisfy
$-1\le\lambda_i\le 1$, which leads to the following asymptotic behavior:
\begin{equation}
\begin{array}{lll}
\lambda=1\;\;\;\;\;\;\Longrightarrow \lambda^\infty=1\\
\lambda=-1\;\;\;\Longrightarrow \lambda^\infty=\pm1\\
\lambda\ne\pm1\;\;\;\Longrightarrow \lambda^\infty=0
\end{array}
\end{equation}
Thus, the time evolution of the state, as dictated by the bi-graded matrix,
will be projected, at the very end, on two eigenvectors corresponding to
the $\pm1$ eigenvalues. The projection on the other eigenvectors will
meanwhile be gradually weakened, due to their eigenvalues approach to zero.
The asymptotic state will thus consist in a superposition of two stationary
states corresponding to the eigenvectors with eigenvalues $\lambda=\pm 1$).
When the initial state of the system precisely coincides with the eigenvector which corresponds to $\lambda=1$ then this state will live forever (like a soliton). If the initial state of the system was exactly the eigenvector
corresponding to $\lambda=-1$, this state will be a stationary wave.\\

Only states which correspond to the $\pm1$ eigenvalue survive
asymptotically (in the even case, this is the situation from the
beginning). During the time evolution the $-1$ eigenvalue 
alternates, the time average for this eigenvalue thus tending to 
zero. In a measurement process with time averaging, the state corresponding
to the $-1$ eigenvalue will thus not be observed. Note, however,
that we identified the eigenvectors with spinors, which would imply that
the spinor and its anti-spinor have different effective eigenvalues under
time averaging. In other words, if the result of the measuring apparatus is
proportional to:
$\left<\psi\right>\;=\;\frac 1{N}\sum_{n=0}^{N}D^n\psi\left(0\right)$\,\,\,
we would obtain the following behavior:\\

$\left<\psi\right>_N\;=\;\left\{
\begin{array}{lll}
\psi\left(0\right) &\,& \lambda=1\\
\\
\pm\frac 1N \psi\left(0\right)$\,\,\,or\,\,\,$ 0 &\,& \lambda=-1
\end{array}\right.$
\\

This is evocative of neutrino phenomenology, in which only the left-chiral
brand is observed, with the right-chiral species perhaps entirely absent. 
One wonders whether such considerations can be applied to quantum field
theory, assuming that the spacetime substratum becomes lattice-like at
Planck distances. We would then have an explanation of the absence of
right-handed neutrinos ``from first principles". \\
 
Another approach would consist in assuming that measurements are only
sensitive to Markovian matrices (which are probabilistic) i.e. only to
the even time steps. In this interpretation all eigenvalues are positive
(with double degeneracy), and one can thus not distinguish between
spinors and anti-spinors. This can be the case, for instance, if the
relevant expectation value is defined as
$\left<\right.\psi\left(t_N\right)\left|\right.\psi\left(t_N\right)\left.\right>\;=\;\left<\right.\psi\left(0\right)\left|\right.\left(D^+\right)^ND^N\left.\right|\psi\left(0\right)\left.\right>$
\,\,\, (where $D^+\;=\;D$).\\

{\large Final remarks}  
Although we have used the term ``bi-graded Markovian" for the
initial matrices only, i.e. before exponentiation, it should be clear
that every odd power of these matrices shares the same property of
becoming Markovian when squared.\\

We suggest that bi-graded Markovian matrices might be applied to a variety
of new quantum and stochastic phenomena. It is thus of much importance to
know more about the mathematics which govern these matrices.\\

As a final remark, we conjecture that it might be possible to construct
$q-$graded Markovian matrices as well. It would then follow that the entries
be complex numbers, which essentially represent complex probabilities.
They might thus be related to noncommutative probability theory \cite{ncp} and to $q-$statistics\footnote{By $q-$statistics we denote particles obeying $q-$commutation relations, defined as:\\ $\left[a\;,\;a^+\right]_q\;=\;aa^+\;-\;qa^+a\;=\;\hbar$ where $-1\;\le\;q\;\le\;1$} \cite{qstat}, as well as to quantum-groups \cite{qg}.\\
 
{\large Acknowledgments}

I would like to thank Prof. Y. Ne'eman and Prof. Z. Schuss for useful
discussions. I would also like to thank Prof. A. Connes for inviting me
to the IHES, where this work was partially written, and the IHES Director,
for the Institute's warm hospitality.
\\ \\ \\
{\LARGE Appendix}
\newline\newline
{\large Appendix A}. Some of the first few bi-graded Markovian matrices in the odd case, and their eigenvalues with the corresponding eigenvectors: 
\newline\newline
$\left( 
\begin{array}{ccc}
0 & \frac 1{\sqrt{2}} & 0 \\ 
\frac 1{\sqrt{2}} & 0 & \frac 1{\sqrt{2}} \\ 
0 & \frac 1{\sqrt{2}} & 0
\end{array}
\right) \;\;\;\;\;\;\;\;\; \Longrightarrow \;\;\;
\allowbreak
\begin{array}{lllll}
\left\{\;\;\; 1\leftrightarrow \left( 
\begin{array}{c}
\frac 12 \;\;,\;\; \frac 1{\sqrt{2}} \;\;,\;\; \frac 12
\end{array}
\right) \right\} \\
\left\{ -1\leftrightarrow
\left( 
\begin{array}{c} -\frac 12 \;\;,\;\; \frac 1{\sqrt{2}} \;\;,\;\; -\frac 12
\end{array}
\right) \right\} \\
\left\{\;\;\; 0\leftrightarrow \left(
\begin{array}{c} -\frac 1{\sqrt{2}} \;\;,\;\; 0 \;\;,\;\; \frac 1{\sqrt{2}}
\end{array}
\right) \right\}
\end{array} $
\\ \\ \\
$\left( 
\begin{array}{ccccc}
0 & \frac 2{\sqrt{6}} & 0 & 0 & 0 \\ 
\frac 2{\sqrt{6}} & 0 & \frac 1{\sqrt{6}} & 0 & 0 \\ 
0 & \frac 1{\sqrt{6}} & 0 & \frac 1{\sqrt{6}} & 0 \\ 
0 & 0 & \frac 1{\sqrt{6}} & 0 & \frac 2{\sqrt{6}} \\ 
0 & 0 & 0 & \frac 2{\sqrt{6}} & 0
\end{array}
\right) \;\;\;\;\;\;\;\;\; \Longrightarrow \;\;\; 
\begin{array}{lllllllll}
\left\{\;\;\; 1\leftrightarrow \left( 
\begin{array}{c}
\frac 1{\sqrt{6}} \;\;,\;\;
\frac 12 \;\;,\;\;   
\frac 1{\sqrt{6}} \;\;,\;\; 
\frac 12 \;\;,\;\; 
\frac 1{\sqrt{6}}
\end{array}
\right) \right\} \\
\left\{ -1\leftrightarrow
\left( 
\begin{array}{c}
\frac 1{\sqrt{6}} \;\;,\;\;
-\frac 12 \;\;,\;\; 
\frac 1{\sqrt{6}}\;\;,\;\; 
-\frac 12 \;\;,\;\;
\frac 1{\sqrt{6}}
\end{array}
\right) \right\} \\
\left\{\;\;\; 0\leftrightarrow \left( 
\begin{array}{c}
\frac 1{\sqrt{6}} \;\;,\;\; 0 \;\;,\;\; -\frac 2{\sqrt{6}} \;\;,\;\; 0 \;\;,\;\; \frac 1{\sqrt{6}}
\end{array}
\right) \right\} \\
\left\{ -\sqrt{\frac 23}\leftrightarrow
\left( 
\begin{array}{c}
-\frac 12 \;\;,\;\;
\frac 12 \;\;,\;\;
0 \;\;,\;\;
-\frac 12 \;\;,\;\;
\frac 12
\end{array}
\right) \right\} \\
\left \{\;\;\; \sqrt{\frac 23}%
\leftrightarrow \left( 
\begin{array}{c}
-\frac 12 \;\;,\;\;
-\frac 12 \;\;,\;\;
0 \;\;,\;\;
\frac 12 \;\;,\;\;
\frac 12
\end{array}
\right) \right\}
\end{array} $ 
\\ \\ \\
$\left( 
\begin{array}{ccccccc}
0 & \frac 3{\sqrt{12}} & 0 & 0 & 0 & 0 & 0 \\ 
\frac 3{\sqrt{12}} & 0 & \frac 1{\sqrt{12}} & 0 & 0 & 0 & 0 \\ 
0 & \frac 1{\sqrt{12}} & 0 & \frac 2{\sqrt{12}} & 0 & 0 & 0 \\ 
0 & 0 & \frac 2{\sqrt{12}} & 0 & \frac 2{\sqrt{12}} & 0 & 0 \\ 
0 & 0 & 0 & \frac 2{\sqrt{12}} & 0 & \frac 1{\sqrt{12}} & 0 \\ 
0 & 0 & 0 & 0 & \frac 1{\sqrt{12}} & 0 & \frac 3{\sqrt{12}} \\ 
0 & 0 & 0 & 0 & 0 & \frac 3{\sqrt{12}} & 0
\end{array}
\right)\;\;\;\;\;\;\;\;\;\Longrightarrow \\ \\ \\
\begin{array}{lllllllllllll}
\left\{\;\;\; 1\leftrightarrow \left( 
\begin{array}{c}
\frac 1{\sqrt{8}} \;\;,\;\;
\frac 1{\sqrt{6}} \;\;,\;\;
\frac 1{\sqrt{8}} \;\;,\;\; 
\frac 1{\sqrt{6}} \;\;,\;\;
\frac 1{\sqrt{8}} \;\;,\;\;
\frac 1{\sqrt{6}} \;\;,\;\;
\frac 1{\sqrt{8}}
\end{array}
\right) \right\} \\
\left\{ -1\leftrightarrow
\left( 
\begin{array}{c}
-\frac 1{\sqrt{8}} \;\;,\;\;  
\frac 1{\sqrt{6}} \;\;,\;\;
-\frac 1{\sqrt{8}} \;\;,\;\;
\frac 1{\sqrt{6}} \;\;,\;\;
-\frac 1{\sqrt{8}} \;\;,\;\;
\frac 1{\sqrt{6}} \;\;,\;\;
-\frac 1{\sqrt{8}}
\end{array}
\right) \right\} \\
\left\{\;\;\; 0\leftrightarrow
\left( 
\begin{array}{c}
-\frac 1{\sqrt{8}} \;\;,\;\;
0 \;\;,\;\;
\frac 3{\sqrt{8}} \;\;,\;\;
0 \;\;,\;\;
-\frac 3{\sqrt{8}} \;\;,\;\;
0 \;\;,\;\;
\frac 1{\sqrt{8}}
\end{array}
\right) \right\} \\
\left\{\;\;\; \sqrt{\frac 56} \leftrightarrow
\left( 
\begin{array}{c}
-\frac 3{\sqrt{40}} \;\;,\;\;
-\frac 12 \;\;,\;\;
-\frac 1{\sqrt{40}} \;\;,\;\;
0 \;\;,\;\;
\frac 1{\sqrt{40}} \;\;,\;\;
\frac 12 \;\;,\;\;
\frac 3{\sqrt{40}}
\end{array}
\right) \right\} \\
\left\{ -\sqrt{\frac 56}%
\leftrightarrow \left( 
\begin{array}{c}
-\frac 3{\sqrt{40}} \;\;,\;\;
\frac 12 \;\;,\;\;
-\frac 1{\sqrt{40}} \;\;,\;\;
0 \;\;,\;\;
\frac 1{\sqrt{40}} \;\;,\;\;
-\frac 12 \;\;,\;\;
\frac 3{\sqrt{40}}
\end{array}
\right) \right\} \\
\left\{\;\;\; \frac 1{\sqrt{2}}\leftrightarrow
\left( 
\begin{array}{c}
-\frac 1{\sqrt{8}} \;\;,\;\;
-\frac 1{\sqrt{12}} \;\;,\;\;
\frac 1{\sqrt{8}} \;\;,\;\;
\frac 1{\sqrt{3}} \;\;,\;\; 
\frac 1{\sqrt{8}} \;\;,\;\; 
-\frac 1{\sqrt{12}} \;\;,\;\;  
-\frac 1{\sqrt{8}}
\end{array}
\right) \right\} \\
\left\{ -\frac 1{\sqrt{2}}%
\leftrightarrow \left( 
\begin{array}{c}
-\frac 1{\sqrt{8}} \;\;,\;\;
\frac 1{\sqrt{12}} \;\;,\;\;
\frac 1{\sqrt{8}} \;\;,\;\;
-\frac 1{\sqrt{3}} \;\;,\;\;
\frac 1{\sqrt{8}} \;\;,\;\;
\frac 1{\sqrt{12}} \;\;,\;\;
-\frac 1{\sqrt{8}}
\end{array}
\right) \right\} 
\end{array} $ 
\newpage
$\left( 
\begin{array}{ccccccccc}
0 & \frac 4{\sqrt{20}} & 0 & 0 & 0 & 0 & 0 & 0 & 0 \\ 
\frac 4{\sqrt{20}} & 0 & \frac 1{\sqrt{20}} & 0 & 0 & 0 & 0 & 0 & 0 \\ 
0 & \frac 1{\sqrt{20}} & 0 & \frac 3{\sqrt{20}} & 0 & 0 & 0 & 0 & 0 \\ 
0 & 0 & \frac 3{\sqrt{20}} & 0 & \frac 2{\sqrt{20}} & 0 & 0 & 0 & 0 \\ 
0 & 0 & 0 & \frac 2{\sqrt{20}} & 0 & \frac 2{\sqrt{20}} & 0 & 0 & 0 \\ 
0 & 0 & 0 & 0 & \frac 2{\sqrt{20}} & 0 & \frac 3{\sqrt{20}} & 0 & 0 \\ 
0 & 0 & 0 & 0 & 0 & \frac 3{\sqrt{20}} & 0 & \frac 1{\sqrt{20}} & 0 \\ 
0 & 0 & 0 & 0 & 0 & 0 & \frac 1{\sqrt{20}} & 0 & \frac 4{\sqrt{20}} \\ 
0 & 0 & 0 & 0 & 0 & 0 & 0 & \frac 4{\sqrt{20}} & 0
\end{array}
\right) \;\;\;\;\;\;\;\Longrightarrow \\ \\ \\ 
\begin{array}{lllllllllllllllll}
\left\{\;\;\; 0\leftrightarrow \left( 
\begin{array}{c}
\frac 1{\sqrt{70}} \;\;,\;\; 
0 \;\;,\;\; 
-\frac 4{\sqrt{70}} \;\;,\;\; 
0 \;\;,\;\; 
\frac 6{\sqrt{70}} \;\;,\;\; 
0 \;\;,\;\; 
-\frac 4{\sqrt{70}} \;\;,\;\; 
0 \;\;,\;\; 
\frac 1{\sqrt{70}} 
\end{array}
\right) \right\} \\ \\
\left\{ -1\leftrightarrow
\left( 
\begin{array}{c}
\frac 1{\sqrt{10}} \;\;,\;\;
-\frac 1{\sqrt{8}} \;\;,\;\;
\frac 1{\sqrt{10}} \;\;,\;\; 
-\frac 1{\sqrt{8}} \;\;,\;\;
\frac 1{\sqrt{10}} \;\;,\;\;
-\frac 1{\sqrt{8}} \;\;,\;\;
\frac 1{\sqrt{10}} \;\;,\;\;
-\frac 1{\sqrt{8}} \;\;,\;\;
\frac 1{\sqrt{10}}
\end{array}
\right) \right\} \\ \\
\left\{\;\;\; 1\leftrightarrow
\left( 
\begin{array}{c}
\frac 1{\sqrt{10}} \;\;,\;\;
\frac 1{\sqrt{8}} \;\;,\;\; 
\frac 1{\sqrt{10}} \;\;,\;\;
\frac 1{\sqrt{8}} \;\;,\;\;
\frac 1{\sqrt{10}} \;\;,\;\;
\frac 1{\sqrt{8}} \;\;,\;\; 
\frac 1{\sqrt{10}} \;\;,\;\;
\frac 1{\sqrt{8}} \;\;,\;\;
\frac 1{\sqrt{10}}
\end{array}
\right) \right\} \\ \\
\left\{ \;\;\; \sqrt {\frac 7{10}}%
\leftrightarrow \left( 
\begin{array}{c}
\frac 1{\sqrt{7}} \;\;,\;\;
\frac 1{\sqrt{80}} \;\;,\;\;
-\frac 1{\sqrt{28}} \;\;,\;\;
-\frac 1{\sqrt{80}} \;\;,\;\;
-\frac 1{\sqrt{7}} \;\;,\;\;
-\frac 1{\sqrt{80}} \;\;,\;\;
-\frac 1{\sqrt{28}} \;\;,\;\;
\frac 1{\sqrt{80}} \;\;,\;\;
\frac 1{\sqrt{7}}
\end{array}
\right) \right\} \\ \\
\left\{ -\sqrt{\frac 7{10}}%
\leftrightarrow \left( 
\begin{array}{c}
\frac 1{\sqrt{7}} \;\;,\;\;
-\frac 1{\sqrt{80}} \;\;,\;\;
-\frac 1{\sqrt{28}} \;\;,\;\;
\frac 1{\sqrt{80}} \;\;,\;\;
-\frac 1{\sqrt{7}} \;\;,\;\;
\frac 1{\sqrt{80}} \;\;,\;\;
-\frac 1{\sqrt{28}} \;\;,\;\;
-\frac 1{\sqrt{80}} \;\;,\;\;
\frac 1{\sqrt{7}}
\end{array}
\right) \right\} \\ \\
\left\{\;\;\;\sqrt{\frac 25} \leftrightarrow
\left( 
\begin{array}{c}
-\frac 1{\sqrt{20}} \;\;,\;\;
-\frac 1{\sqrt{40}} \;\;,\;\;
\frac 2{\sqrt{20}} \;\;,\;\;
\frac 3{\sqrt{40}} \;\;,\;\;
0 \;\;,\;\;
-\frac 3{\sqrt{40}} \;\;,\;\;
-\frac 2{\sqrt{20}} \;\;,\;\;
\frac 1{\sqrt{40}} \;\;,\;\;
\frac 1{\sqrt{20}}
\end{array}
\right) \right\} \\ \\
\left\{ -\sqrt{\frac 25}
\leftrightarrow \left( 
\begin{array}{c}
\frac 1{\sqrt{20}} \;\;,\;\;
-\frac 1{\sqrt{40}} \;\;,\;\;
-\frac 2{\sqrt{20}} \;\;,\;\;
\frac 3{\sqrt{40}} \;\;,\;\;
0 \;\;,\;\;
-\frac 3{\sqrt{40}} \;\;,\;\;
\frac 2{\sqrt{20}} \;\;,\;\;
\frac 1{\sqrt{40}} \;\;,\;\;
-\frac 1{\sqrt{20}}
\end{array}
\right) \right\} \\ \\
\left\{\;\;\; \frac 3{\sqrt{10}}%
\leftrightarrow \left( 
\begin{array}{c}
\frac 1{\sqrt{5}} \;\;,\;\;
\frac 3{\sqrt{40}} \;\;,\;\;
\frac 1{\sqrt{20}} \;\;,\;\;
\frac 1{\sqrt{40}} \;\;,\;\;
0 \;\;,\;\;
-\frac 1{\sqrt{40}} \;\;,\;\;
-\frac 1{\sqrt{20}} \;\;,\;\;
-\frac 3{\sqrt{40}} \;\;,\;\;
-\frac 1{\sqrt{5}}
\end{array}
\right) \right\} \\ \\
\left\{ -\frac 3{\sqrt{10}}%
\leftrightarrow \left( 
\begin{array}{c}
\frac 1{\sqrt{5}} \;\;,\;\;
-\frac 3{\sqrt{40}} \;\;,\;\;
\frac 1{\sqrt{20}} \;\;,\;\;
-\frac 1{\sqrt{40}} \;\;,\;\;
0 \;\;,\;\;
\frac 1{\sqrt{40}} \;\;,\;\;
-\frac 1{\sqrt{20}} \;\;,\;\;
\frac 3{\sqrt{40}} \;\;,\;\;
-\frac 1{\sqrt{5}}
\end{array}
\right) \right\} 
\end{array}$

\newpage

{\large Appendix B.}  Some of the first few bi-graded Markovian matrices in the even case, and their eigenvalues with the corresponding eigenvectors:
\newline\newline\newline
$\left( 
\begin{array}{cc}
0 & 1 \\ 
1 & 0
\end{array}
\right)\;\;\;\;
\Longrightarrow
\left\{ 1\leftrightarrow \left( 1\;,\;1\right) \right\} ,\allowbreak 
\left\{ -1\leftrightarrow \left( -1\;,\;1\right) \right\}$
\newline
\newline
\newline
\newline
$\left( 
\begin{array}{cccc}
0 & 1 & 0 & 0 \\ 
1 & 0 & 0 & 0 \\ 
0 & 0 & 0 & 1 \\ 
0 & 0 & 1 & 0
\end{array}
\right)\;\;\;\Longrightarrow
\begin{array}{lll}
\left\{ 1\leftrightarrow \left( 1\;,\;1\;,\; 0\;,\; 0\right)\right\}\\
\\
\left\{ 1\leftrightarrow \left( 0\;,\;0\;,\;1\;,\;1\right) \right\} 
\end{array}$
\allowbreak
$\begin{array}{lll}
\left\{ -1\leftrightarrow \left( -1\;,\;1\;,\;0\;,\;0\right)\right\}\\
\\
\left\{ -1\leftrightarrow \left( 0\;,\;0\;,\;-1\;,\;1\right) \right\}
\end{array}$
\newline
\newline
\newline
\newline
$\left( 
\begin{array}{cccccc}
0 & 1 & 0 & 0 & 0 & 0 \\ 
1 & 0 & 0 & 0 & 0 & 0 \\ 
0 & 0 & 0 & 1 & 0 & 0 \\ 
0 & 0 & 1 & 0 & 0 & 0 \\ 
0 & 0 & 0 & 0 & 0 & 1 \\ 
0 & 0 & 0 & 0 & 1 & 0
\end{array}
\right)\;\;\;\;
\Longrightarrow
\\ \\ \\
\begin{array}{lllll}
\left\{ 1\leftrightarrow \left( 1 \;,\;1 \;,\; 0\;,\; 0 \;,\;0\;,\;0\right)\right\}\\
 \\
\left\{ 1\leftrightarrow \left( 0 \;,\;0 \;,\;1 \;,\;1 \;,\;0
    \;,\;0\right)\right\}\\ 
\\
\left\{ 1\leftrightarrow\left( 0 \;,\;0 \;,\;0 \;,\;0 \;,\;1 \;,\;
    1\right) \right\}
\end{array}$ 
\allowbreak 
$\begin{array}{lllll}
\left\{ -1\leftrightarrow \left( -1 \;,\; 1 \;,\;0 \;,\;0 \;,\;0
    \;,\;0\right)\right\}\\
\\
\left\{ -1\leftrightarrow\left( 0 \;,\;0 \;,\;-1 \;,\; 1 \;,\;0 \;,\;
    0\right)\right\}\\
\\
\left\{ -1\leftrightarrow\left( 0 \;,\;0 \;,\;0 \;,\;0 \;,\;-1
    \;,\;1\right) \right\}
\end{array} $
\newpage
{\large Appendix C.} An example for a bi-graded matrix evolution in the odd case:
\newline\newline\newline
$A \; = \;\left( 
\begin{array}{ccccccccc}
0 & \frac 4{\sqrt{20}} & 0 & 0 & 0 & 0 & 0 & 0 & 0 \\ 
\frac 4{\sqrt{20}} & 0 & \frac 1{\sqrt{20}} & 0 & 0 & 0 & 0 & 0 & 0 \\ 
0 & \frac 1{\sqrt{20}} & 0 & \frac 3{\sqrt{20}} & 0 & 0 & 0 & 0 & 0 \\ 
0 & 0 & \frac 3{\sqrt{20}} & 0 & \frac 2{\sqrt{20}} & 0 & 0 & 0 & 0 \\ 
0 & 0 & 0 & \frac 2{\sqrt{20}} & 0 & \frac 2{\sqrt{20}} & 0 & 0 & 0 \\ 
0 & 0 & 0 & 0 & \frac 2{\sqrt{20}} & 0 & \frac 3{\sqrt{20}} & 0 & 0 \\ 
0 & 0 & 0 & 0 & 0 & \frac 3{\sqrt{20}} & 0 & \frac 1{\sqrt{20}} & 0 \\ 
0 & 0 & 0 & 0 & 0 & 0 & \frac 1{\sqrt{20}} & 0 & \frac 4{\sqrt{20}} \\ 
0 & 0 & 0 & 0 & 0 & 0 & 0 & \frac 4{\sqrt{20}} & 0
\end{array}
\right)$
\newline\newline\newline
$A^2=\left( 
\begin{array}{ccccccccc}
.8 & 0 & .2 & 0 & 0 & 0 & 0 & 0 & 0 \\ 
0 & .85 & 0 & .15 & 0 & 0 & 0 & 0 & 0 \\ 
.2 & 0 & .5 & 0 & .3 & 0 & 0 & 0 & 0 \\ 
0 & .15 & 0 & .65 & 0 & .2 & 0 & 0 & 0 \\ 
0 & 0 & .3 & 0 & .4 & 0 & .3 & 0 & 0 \\ 
0 & 0 & 0 & .2 & 0 & .65 & 0 & .15 & 0 \\ 
0 & 0 & 0 & 0 & .3 & 0 & .5 & 0 & .2 \\ 
0 & 0 & 0 & 0 & 0 & .15 & 0 & .85 & 0 \\ 
0 & 0 & 0 & 0 & 0 & 0 & .2 & 0 & .8
\end{array}
\right) $
\newline
\newline
\newline
$A^3=\left( 
\begin{array}{ccccccccc}
0 & .76026 & 0 & .13416 & 0 & 0 & 0 & 0 & 0 \\ 
.76026 & 0 & .29069 & 0 & .06708 & 0 & 0 & 0 & 0 \\ 
0 & .29069 & 0 & .46957 & 0 & .13416 & 0 & 0 & 0 \\ 
.13416 & 0 & .46957 & 0 & .38013 & 0 & .13416 & 0 & 0 \\ 
0 & .06708 & 0 & .38013 & 0 & .38013 & 0 & .06708
& 0 \\ 
0 & 0 & .13416 & 0 & .38013 & 0 & .46957 & 0 & .13416 \\ 
0 & 0 & 0 & .13416 & 0 & .46957 & 0 & .29069 & 0 \\ 
0 & 0 & 0 & 0 & .06708 & 0 & .29069 & 0 & .76026 \\ 
0 & 0 & 0 & 0 & 0 & .13416 & 0 & .76026 & 0
\end{array}
\right) $
\newline\newline\newline
$A^4=\left( 
\begin{array}{ccccccccc}
.68 & 0 & .26 & 0 & .06 & 0 & 0 & 0 & 0 \\ 
0 & .745 & 0 & .225 & 0 & .03 & 0 & 0 & 0 \\ 
.26 & 0 & .38 & 0 & .27 & 0 & .09 & 0 & 0 \\ 
0 & .225 & 0 & .485 & 0 & .26 & 0 & .03 & 0 \\ 
.06 & 0 & .27 & 0 & .34 & 0 & .27 & 0 & .06 \\ 
0 & .03 & 0 & .26 & 0 & .485 & 0 & .225 & 0 \\ 
0 & 0 & .09 & 0 & .27 & 0 & .38 & 0 & .26 \\ 
0 & 0 & 0 & .03 & 0 & .225 & 0 & .745 & 0 \\ 
0 & 0 & 0 & 0 & .06 & 0 & .26 & 0 & .68
\end{array}
\right) $
\newline\newline\newline
$A^7=\left( 
\begin{array}{ccccccccc}
0 & .59658 & 0 & .23613 & 0 & .05769 & 0 & .00405
& 0 \\ 
.59658 & 0 & .32624 & 0 & .14691 & 0 & .04427 & 0 & 
.00405 \\ 
0 & .32624 & 0 & .32803 & 0 & .19588 & 0 & .04427 & 0 \\ 
.23613 & 0 & .32803 & 0 & .3003 & 0 & .19588 & 0 & .05769 \\ 
0 & .14691 & 0 & .3003 & 0 & .3003 & 0 & .14691 & 0 \\ 
.05769 & 0 & .19588 & 0 & .3003 & 0 & .32803 & 0 & .23613 \\ 
0 & .04427 & 0 & .19588 & 0 & .32803 & 0 & .32624 & 0 \\ 
.00405 & 0 & .04427 & 0 & .14691 & 0 & .32624 & 0
& .59658 \\ 
0 & .00405 & 0 & .05769 & 0 & .23613 & 0 & .59658
& 0
\end{array}
\right) $
\newline\newline\newline
$A^8=\left( 
\begin{array}{ccccccccc}
.5336 & 0 & .2918 & 0 & .1314 & 0 & .0396 & 0 & .0036 \\ 
0 & .60655 & 0 & .28455 & 0 & .0954 & 0 & .0135 & 0 \\ 
.2918 & 0 & .293 & 0 & .2343 & 0 & .1413 & 0 & .0396 \\ 
0 & .28455 & 0 & .35435 & 0 & .2657 & 0 & .0954 & 0 \\ 
.1314 & 0 & .2343 & 0 & .2686 & 0 & .2343 & 0 & .1314 \\ 
0 & .0954 & 0 & .2657 & 0 & .35435 & 0 & .28455 & 0 \\ 
.0396 & 0 & .1413 & 0 & .2343 & 0 & .293 & 0 & .2918 \\ 
0 & .0135 & 0 & .0954 & 0 & .28455 & 0 & .60655 & 0 \\ 
.0036 & 0 & .0396 & 0 & .1314 & 0 & .2918 & 0 & .5336
\end{array}
\right) $
\newline\newline\newline
$A^{15}=\left( 
\begin{array}{ccccccccc}
0 & .43461 & 0 & .26914 & 0 & .14124 & 0 & .04944 & 0 \\ 
.43461 & 0 & .31051 & 0 & .20519 & 0 & .11829 & 0 & .04944 \\ 
0 & .31051 & 0 & .26534 & 0 & .20029 & 0 & .11829 & 0 \\ 
.26914 & 0 & .26534 & 0 & .24202 & 0 & .20029 & 0 & .14124 \\ 
0 & .20519 & 0 & .24202 & 0 & .24202 & 0 & .20519 & 0 \\ 
.14124 & 0 & .20029 & 0 & .24202 & 0 & .26534 & 0 & .26914 \\ 
0 & .11829 & 0 & .20029 & 0 & .26534 & 0 & .31051 & 0 \\ 
.04944 & 0 & .11829 & 0 & .20519 & 0 & .31051 & 0 & .43461 \\ 
0 & .04944 & 0 & .14124 & 0 & .26914 & 0 & .43461 & 0
\end{array}
\right) $
\newline\newline\newline
$A^{16}=\left( 
\begin{array}{ccccccccc}
.38872 & 0 & .27773 & 0 & .18353 & 0 & .1058 & 0 & .04422 \\ 
0 & .45816 & 0 & .30006 & 0 & .17112 & 0 & .07067 & 0 \\ 
.27773 & 0 & .24743 & 0 & .20824 & 0 & .16081 & 0 & .1058 \\ 
0 & .30006 & 0 & .28623 & 0 & .24259 & 0 & .17112 & 0 \\ 
.18353 & 0 & .20824 & 0 & .21647 & 0 & .20824 & 0 & .18353 \\ 
0 & .17112 & 0 & .24259 & 0 & .28623 & 0 & .30006 & 0 \\ 
.1058 & 0 & .16081 & 0 & .20824 & 0 & .24743 & 0 & .27773 \\ 
0 & .07067 & 0 & .17112 & 0 & .30006 & 0 & .45816 & 0 \\ 
.04422 & 0 & .1058 & 0 & .18353 & 0 & .27773 & 0 & .38872
\end{array}
\right) $
\newline\newline\newline
$A^{31}=\left( 
\begin{array}{ccccccccc}
0 & .30754 & 0 & .25017 & 0 & .19492 & 0 & .1418 & 0 \\ 
.30754 & 0 & .26451 & 0 & .22255 & 0 & .18164 & 0 & .1418 \\ 
0 & .26451 & 0 & .23795 & 0 & .21033 & 0 & .18164 & 0 \\ 
.25017 & 0 & .23795 & 0 & .22467 & 0 & .21033 & 0 & .19492 \\ 
0 & .22255 & 0 & .22467 & 0 & .22467 & 0 & .22255 & 0 \\ 
.19492 & 0 & .21033 & 0 & .22467 & 0 & .23795 & 0 & .25017 \\ 
0 & .18164 & 0 & .21033 & 0 & .23795 & 0 & .26451 & 0 \\ 
.1418 & 0 & .18164 & 0 & .22255 & 0 & .26451 & 0 & .30754 \\ 
0 & .1418 & 0 & .19492 & 0 & .25017 & 0 & .30754 & 0
\end{array}
\right) $
\newline\newline\newline
$A^{32}=\left( 
\begin{array}{ccccccccc}
.27507 & 0 & .23659 & 0 & .19905 & 0 & .16246 & 0 & .12683 \\ 
0 & .33422 & 0 & .27696 & 0 & .22137 & 0 & .16744 & 0 \\ 
.23659 & 0 & .21877 & 0 & .20047 & 0 & .18171 & 0 & .16246 \\ 
0 & .27696 & 0 & .2601 & 0 & .24157 & 0 & .22137 & 0 \\ 
.19905 & 0 & .20047 & 0 & .20095 & 0 & .20047 & 0 & .19905 \\ 
0 & .22137 & 0 & .24157 & 0 & .2601 & 0 & .27696 & 0 \\ 
.16246 & 0 & .18171 & 0 & .20047 & 0 & .21877 & 0 & .23659 \\ 
0 & .16744 & 0 & .22137 & 0 & .27696 & 0 & .33422 & 0 \\ 
.12683 & 0 & .16246 & 0 & .19905 & 0 & .23659 & 0 & .27507
\end{array}
\right) $
\newline\newline\newline
$A^{63}=\left( 
\begin{array}{ccccccccc}
0 & .23897 & 0 & .22872 & 0 & .21848 & 0 & .20825 & 0 \\ 
.23897 & 0 & .23128 & 0 & .2236 & 0 & .21593 & 0 & .20825 \\ 
0 & .23128 & 0 & .22617 & 0 & .22105 & 0 & .21593 & 0 \\ 
.22872 & 0 & .22617 & 0 & .22361 & 0 & .22105 & 0 & .21848 \\ 
0 & .2236 & 0 & .22361 & 0 & .22361 & 0 & .2236 & 0 \\ 
.21848 & 0 & .22105 & 0 & .22361 & 0 & .22617 & 0 & .22872 \\ 
0 & .21593 & 0 & .22105 & 0 & .22617 & 0 & .23128 & 0 \\ 
.20825 & 0 & .21593 & 0 & .2236 & 0 & .23128 & 0 & .23897 \\ 
0 & .20825 & 0 & .21848 & 0 & .22872 & 0 & .23897 & 0
\end{array}
\right) $
\newline\newline\newline
$A^{64}=\left( 
\begin{array}{ccccccccc}
.21374 & 0 & .20687 & 0 & .2 & 0 & .19313 & 0 & .18627 \\
0 & .26545 & 0 & .25515 & 0 & .24485 & 0 & .23455 & 0 \\
.20687 & 0 & .20343 & 0 & .2 & 0 & .19657 & 0 & .19313 \\
0 & .25515 & 0 & .25172 & 0 & .24829 & 0 & .24485 & 0 \\
.2 & 0 & .2 & 0 & .2 & 0 & .2 & 0 & .2 \\
0 & .24485 & 0 & .24829 & 0 & .25172 & 0 & .25515 & 0 \\
.19313 & 0 & .19657 & 0 & .2 & 0 & .20343 & 0 & .20687 \\
0 & .23455 & 0 & .24485 & 0 & .25515 & 0 & .26545 & 0 \\
.18627 & 0 & .19313 & 0 & .2 & 0 & .20687 & 0 & .21374 \\
\end{array}
\right) $
\newline\newline\newline
$A^{127}=\left( 
\begin{array}{ccccccccc}
0 & .22413 & 0 & .22378 & 0 & .22343 & 0 & .22308 & 0 \\ 
.22413 & 0 & .22387 & 0 & .22361 & 0 & .22334 & 0 & .22308 \\ 
0 & .22387 & 0 & .22369 & 0 & .22352 & 0 & .22334 & 0 \\ 
.22378 & 0 & .22369 & 0 & .22361 & 0 & .22352 & 0 & .22343 \\ 
0 & .22361 & 0 & .22361 & 0 & .22361 & 0 & .22361 & 0 \\ 
.22343 & 0 & .22352 & 0 & .22361 & 0 & .22369 & 0 & .22378 \\ 
0 & .22334 & 0 & .22352 & 0 & .22369 & 0 & .22387 & 0 \\ 
.22308 & 0 & .22334 & 0 & .22361 & 0 & .22387 & 0 & .22413 \\ 
0 & .22308 & 0 & .22343 & 0 & .22378 & 0 & .22413 & 0
\end{array}
\right) $
\newline\newline\newline
$A^{128}=\left( 
\begin{array}{ccccccccc}
.20047 & 0 & .20024 & 0 & .2 & 0 & .19976 & 0 & .19953 \\ 
0 & .25053 & 0 & .25018 & 0 & .24982 & 0 & .24947 & 0 \\ 
.20024 & 0 & .20012 & 0 & .2 & 0 & .19988 & 0 & .19976 \\ 
0 & .25018 & 0 & .25006 & 0 & .24994 & 0 & .24982 & 0 \\ 
.2 & 0 & .2 & 0 & .2 & 0 & .2 & 0 & .2 \\ 
0 & .24982 & 0 & .24994 & 0 & .25006 & 0 & .25018 & 0 \\ 
.19976 & 0 & .19988 & 0 & .2 & 0 & .20012 & 0 & .20024 \\ 
0 & .24947 & 0 & .24982 & 0 & .25018 & 0 & .25053 & 0 \\ 
.19953 & 0 & .19976 & 0 & .2 & 0 & .20024 & 0 & .20047
\end{array}
\right) $
\newline\newline\newline
$A^{255}=\left( 
\begin{array}{ccccccccc}
0 & .22361 & 0 & .22361 & 0 & .22361 & 0 & .22361 & 0 \\ 
.22361 & 0 & .22361 & 0 & .22361 & 0 & .22361 & 0 & .22361 \\ 
0 & .22361 & 0 & .22361 & 0 & .22361 & 0 & .22361 & 0 \\ 
.22361 & 0 & .22361 & 0 & .22361 & 0 & .22361 & 0 & .22361 \\ 
0 & .22361 & 0 & .22361 & 0 & .22361 & 0 & .22361 & 0 \\ 
.22361 & 0 & .22361 & 0 & .22361 & 0 & .22361 & 0 & .22361 \\ 
0 & .22361 & 0 & .22361 & 0 & .22361 & 0 & .22361 & 0 \\ 
.22361 & 0 & .22361 & 0 & .22361 & 0 & .22361 & 0 & .22361 \\ 
0 & .22361 & 0 & .22361 & 0 & .22361 & 0 & .22361 & 0
\end{array}
\right) $
\newline\newline\newline
$A^{256}=\left( 
\begin{array}{ccccccccc}
.2 & 0 & .2 & 0 & .2 & 0 & .2 & 0 & .2 \\ 
0 & .25 & 0 & .25 & 0 & .25 & 0 & .25 & 0 \\ 
.2 & 0 & .2 & 0 & .2 & 0 & .2 & 0 & .2 \\ 
0 & .25 & 0 & .25 & 0 & .25 & 0 & .25 & 0 \\ 
.2 & 0 & .2 & 0 & .2 & 0 & .2 & 0 & .2 \\ 
0 & .25 & 0 & .25 & 0 & .25 & 0 & .25 & 0 \\ 
.2 & 0 & .2 & 0 & .2 & 0 & .2 & 0 & .2 \\ 
0 & .25 & 0 & .25 & 0 & .25 & 0 & .25 & 0 \\ 
.2 & 0 & .2 & 0 & .2 & 0 & .2 & 0 & .2
\end{array}
\right) $\\
\newpage

{\large Appendix D.} The Algebra which governs $M_+$ and $M_-:$\\ 
\\
\,

$M_{+/-}$ are the raising and lowering operators, which together build
the bi-graded Markovian matrices. $Q\,\,\,\& \,\,\,R$ are matrices. The
entries in the $R$ matrix which are denoted by $a_{i,j}$ are free.
The $Q$ matrix is diagonal and rational.\\
\\
\,

$M_{2+}=\left(
\begin{array}{cc}
0 & 1 \\
0 & 0
\end{array}
\right) \;\;\;\;\;M_{2-}=\left(M_{2+}\right)^{tr}=\left(
\begin{array}{cc}
0 & 0 \\
1 & 0
\end{array}
\right) $\\
\\
\,

$M_{2+}M_{2-}-M_{2-}M_{2+}=\left(
\begin{array}{cc}
1 & 0 \\
0 & -1
\end{array}
\right)$\\
\,

$M_{2+}M_{2-}+M_{2-}M_{2+}={\Bbb I}_2$\\
\\
\\
\\
\,

$M_{4+}=\left(
\begin{array}{cccc}
0 & 1 & 0 & 0 \\
0 & 0 & 0 & 0 \\
0 & 0 & 0 & 1 \\
0 & 0 & 0 & 0
\end{array}\right)\;\;\;\;\;\;M_{4-}=\left(M_{4+}\right)^{tr}=\left( 
\begin{array}{cccc}
0 & 0 & 0 & 0 \\ 
1 & 0 & 0 & 0 \\ 
0 & 0 & 0 & 0 \\ 
0 & 0 & 1 & 0
\end{array}
\right) $\\
\\
\,

$M_{4+}M_{4-}-M_{4-}M_{4+}=\left( 
\begin{array}{cccc}
1 & 0 & 0 & 0 \\ 
0 & -1 & 0 & 0 \\ 
0 & 0 & 1 & 0 \\ 
0 & 0 & 0 & -1
\end{array}
\right)$\\
\,

$M_{4+}M_{4-}+M_{4-}M_{4+}={\Bbb I}_4$\\
\\
\\
\\
\,

$M_{6+}=\left( 
\begin{array}{cccccc}
0 & 1 & 0 & 0 & 0 & 0 \\ 
0 & 0 & 0 & 0 & 0 & 0 \\ 
0 & 0 & 0 & 1 & 0 & 0 \\ 
0 & 0 & 0 & 0 & 0 & 0 \\ 
0 & 0 & 0 & 0 & 0 & 1 \\ 
0 & 0 & 0 & 0 & 0 & 0
\end{array}
\right) \;\;\;\;\;\;\;\;M_{6-}=\left(M_{6+}\right)^{tr}$\\
\\
\,

$M_{6+}M_{6-}-M_{6-}M_{6+}=\left( 
\begin{array}{cccccc}
1 & 0 & 0 & 0 & 0 & 0 \\ 
0 & -1 & 0 & 0 & 0 & 0 \\ 
0 & 0 & 1 & 0 & 0 & 0 \\ 
0 & 0 & 0 & -1 & 0 & 0 \\ 
0 & 0 & 0 & 0 & 1 & 0 \\ 
0 & 0 & 0 & 0 & 0 & -1
\end{array}
\right) $\\
\\
\,

$\;M_{6+}M_{6-}+M_{6-}M_{6+}={\Bbb I}_6$\\
\\
\\
\,

$M_{3+}=\left( 
\begin{array}{ccc}
0 & \frac 1{\sqrt{2}} & 0 \\ 
0 & 0 & \frac 1{\sqrt{2}} \\ 
0 & 0 & 0
\end{array}
\right) \;\;\;\;\;\;\;M_{3-}=\left(M_{3+}\right)^{tr} $\\
\\
\,

$R_3=\left( 
\begin{array}{ccc}
0 & 0 & a_{1,3} \\ 
a_{2,1} & a_{2,2} & 0 \\ 
a_{3,1} & a_{3,2} & 0
\end{array}
\right) \;\;\;\;\;\Longrightarrow \;\;\;\;\;\;\;R_{3}M_{3+}M_{3-}=M_{3-}M_{3+}R_{3}$\\
\\
\,

$M_{3+}M_{3-}-M_{3-}M_{3+}=\left( 
\begin{array}{ccc}
\frac 12 & 0 & 0 \\ 
0 & 0 & 0 \\ 
0 & 0 & -\frac 12
\end{array}
\right)$\\
\\
\,

$M_{3+}M_{3-}+M_{3-}M_{3+}=\left( 
\begin{array}{ccc}
\frac 14 & 0 & 0 \\ 
0 & \frac 12 & 0 \\ 
0 & 0 & \frac 12
\end{array}
\right) $\\
\\
\,

$Q_{3}=\left( 
\begin{array}{ccc}
0 & 0 & 0 \\ 
0 & 0 & 0 \\ 
0 & 0 & 1
\end{array}
\right) \;\;\;\;\Longrightarrow
\;\;\;\;M_{3+}M_{3-}+Q_{3}M_{3-}M_{3+}=\frac 12{\Bbb I}_{3}$\\
\\
\\
\\
\,

$M_{5+}=\left( 
\begin{array}{ccccc}
0 & \frac 2{\sqrt{6}} & 0 & 0 & 0 \\
0 & 0 & \frac 1{\sqrt{6}} & 0 & 0 \\
0 & 0 & 0 & \frac 1{\sqrt{6}} & 0 \\
0 & 0 & 0 & 0 & \frac 2{\sqrt{6}} \\
0 & 0 & 0 & 0 & 0
\end{array}
\right) \;\;\;\;\;\;\;\;\;\;\;M_{5-}=\left(M_{5+}\right)^{tr} $\\
\\
\,

$R_{5}=\left( 
\begin{array}{ccccc}
0 & 0 & 0 & 0 & a_{1,5} \\ 
a_{2,1} & 0 & 0 & a_{2,4} & 0 \\
0 & a_{3,2} & a_{3,3} & 0 & 0 \\
0 & a_{4,2} & a_{4,3} & 0 & 0 \\
a_{5,1} & 0 & 0 & a_{5,4} & 0
\end{array}
\right) \;\;\;\;\Longrightarrow \;\;\;\;R_{5}M_{5+}M_{5-}=M_{5-}M_{5+}R_{5}$\\
\\
\,

$M_{5+}M_{5-}-M_{5-}M_{5+}=\left( 
\begin{array}{ccccc}
\frac 23 & 0 & 0 & 0 & 0 \\
0 & -\frac 12 & 0 & 0 & 0 \\
0 & 0 & 0 & 0 & 0 \\
0 & 0 & 0 & \frac 12 & 0 \\
0 & 0 & 0 & 0 & -\frac 23
\end{array}
\right)$\\
\\
\,

$M_{5+}M_{5-}+M_{5-}M_{5+}=\left( 
\begin{array}{ccccc}
\frac 49 & 0 & 0 & 0 & 0 \\
0 & \frac 7{12} & 0 & 0 & 0 \\
0 & 0 & \frac 16 & 0 & 0 \\
0 & 0 & 0 & \frac 12 & 0 \\
0 & 0 & 0 & 0 & \frac 23
\end{array}
\right) $\\
\\
\,

$Q_{5}=\left( 
\begin{array}{ccccc}
0 & 0 & 0 & 0 & 0 \\
0 & \frac 34 & 0 & 0 & 0 \\
0 & 0 & 3 & 0 & 0 \\
0 & 0 & 0 & 0 & 0 \\
0 & 0 & 0 & 0 & 1
\end{array}
\right) \;\;\;\;\Longrightarrow
\;\;\;\;\;M_{5+}M_{5-}+Q_{5}M_{5-}M_{5+}=\frac 23{\Bbb I}_{5}
 $\\
\\
\\
\,

$M_{7+}=\left( 
\begin{array}{ccccccc}
0 & \frac 3{\sqrt{12}} & 0 & 0 & 0 & 0 & 0 \\
0 & 0 & \frac 1{\sqrt{12}} & 0 & 0 & 0 & 0 \\
0 & 0 & 0 & \frac 2{\sqrt{12}} & 0 & 0 & 0 \\
0 & 0 & 0 & 0 & \frac 2{\sqrt{12}} & 0 & 0 \\
0 & 0 & 0 & 0 & 0 & \frac 1{\sqrt{12}} & 0 \\
0 & 0 & 0 & 0 & 0 & 0 & \frac 3{\sqrt{12}} \\
0 & 0 & 0 & 0 & 0 & 0 & 0
\end{array}
\right) \;\;\;\;\;\;M_{7-}=\left(M_{7+}\right)^{tr}$
\\
\\
\,

$R_{7}=\left( 
\begin{array}{ccccccc}
0 & 0 & 0 & 0 & 0 & 0 & a_{1,7} \\
a_{2,1} & 0 & 0 & 0 & 0 & a_{2,6} & 0 \\
0 & a_{3,2} & 0 & 0 & a_{3,5} & 0 & 0 \\
0 & 0 & a_{4,3} & a_{4,4} & 0 & 0 & 0 \\
0 & 0 & a_{5,3} & a_{5,4} & 0 & 0 & 0 \\
0 & a_{6,2} & 0 & 0 & a_{6,5} & 0 & 0 \\
a_{7,1} & 0 & 0 & 0 & 0 & a_{7,6} & 0
\end{array}
\right) \;\;\;\;\Longrightarrow $\\
\\
\,

$\Longrightarrow \;\;\;\; R_{7}M_{7+}M_{7-}=M_{7-}M_{7+}R_{7}$\\
\\
\\
\,

$M_{7+}M_{7-}-M_{7-}M_{7+}=\left( 
\begin{array}{ccccccc}
\frac 34 & 0 & 0 & 0 & 0 & 0 & 0 \\
0 & -\frac 23 & 0 & 0 & 0 & 0 & 0 \\
0 & 0 & \frac 14 & 0 & 0 & 0 & 0 \\
0 & 0 & 0 & 0 & 0 & 0 & 0 \\
0 & 0 & 0 & 0 & -\frac 14 & 0 & 0 \\
0 & 0 & 0 & 0 & 0 & \frac 23 & 0 \\
0 & 0 & 0 & 0 & 0 & 0 & -\frac 34
\end{array}
\right) $\\
\\
\,

$M_{7+}M_{7-}+M_{7-}M_{7+}=\left( 
\begin{array}{ccccccc}
\frac 34 & 0 & 0 & 0 & 0 & 0 & 0 \\
0 & \frac 56 & 0 & 0 & 0 & 0 & 0 \\
0 & 0 & \frac 5{12} & 0 & 0 & 0 & 0 \\
0 & 0 & 0 & \frac 23 & 0 & 0 & 0 \\
0 & 0 & 0 & 0 & \frac 5{12} & 0 & 0 \\
0 & 0 & 0 & 0 & 0 & \frac 56 & 0 \\
0 & 0 & 0 & 0 & 0 & 0 & \frac 34
\end{array}
\right) $\\
\\
\,

$Q_{7}=\left( 
\begin{array}{ccccccc}
0 & 0 & 0 & 0 & 0 & 0 & 0 \\
0 & \frac 89 & 0 & 0 & 0 & 0 & 0 \\
0 & 0 & 5 & 0 & 0 & 0 & 0 \\
0 & 0 & 0 & \frac 54 & 0 & 0 & 0 \\
0 & 0 & 0 & 0 & 2 & 0 & 0 \\
0 & 0 & 0 & 0 & 0 & 0 & 0 \\
0 & 0 & 0 & 0 & 0 & 0 & 1
\end{array}
\right) \;\;\;\;\Longrightarrow
\;\;\;\;\;M_{7+}M_{7-}+Q_{7}M_{7-}M_{7+}=\frac 34{\Bbb I}_{7}$\\
\\
\\
\\
\,

$M_{9+}=\left( 
\begin{array}{ccccccccc}
0 & \frac 4{\sqrt{20}} & 0 & 0 & 0 & 0 & 0 & 0 & 0 \\
0 & 0 & \frac 1{\sqrt{20}} & 0 & 0 & 0 & 0 & 0 & 0 \\
0 & 0 & 0 & \frac 3{\sqrt{20}} & 0 & 0 & 0 & 0 & 0 \\
0 & 0 & 0 & 0 & \frac 2{\sqrt{20}} & 0 & 0 & 0 & 0 \\
0 & 0 & 0 & 0 & 0 & \frac 2{\sqrt{20}} & 0 & 0 & 0 \\
0 & 0 & 0 & 0 & 0 & 0 & \frac 3{\sqrt{20}} & 0 & 0 \\
0 & 0 & 0 & 0 & 0 & 0 & 0 & \frac 1{\sqrt{20}} & 0 \\
0 & 0 & 0 & 0 & 0 & 0 & 0 & 0 & \frac 4{\sqrt{20}} \\
0 & 0 & 0 & 0 & 0 & 0 & 0 & 0 & 0
\end{array}
\right) \;\;\;\;\;\;M_{9-}=\left(M_{9+}\right)^{tr}$\\
\\
\,

$R_{9}=\left( 
\begin{array}{ccccccccc}
0 & 0 & 0 & 0 & 0 & 0 & 0 & 0 & a_{1,9} \\
a_{2,1} & 0 & 0 & 0 & 0 & 0 & 0 & a_{2,8} & 0 \\
0 & a_{3,2} & 0 & 0 & 0 & 0 & a_{3,7} & 0 & 0 \\
0 & 0 & a_{4,3} & 0 & 0 & a_{4,6} & 0 & 0 & 0 \\
0 & 0 & 0 & a_{5,4} & a_{5,5} & 0 & 0 & 0 & 0 \\
0 & 0 & 0 & a_{6,4} & a_{6,5} & 0 & 0 & 0 & 0 \\
0 & 0 & a_{7,3} & 0 & 0 & a_{7,6} & 0 & 0 & 0 \\
0 & a_{8,2} & 0 & 0 & 0 & 0 & a_{8,7} & 0 & 0 \\
a_{9,1} & 0 & 0 & 0 & 0 & 0 & 0 & a_{9,8} & 0
\end{array}
\right) \;\Longrightarrow \;$\\
\\
\,

$\Longrightarrow \;\;R_{9}M_{9+}M_{9-}=M_{9-}M_{9+}R_{9}$\\
\\
\,

$M_{9+}M_{9-}-M_{9-}M_{9+}=\left( 
\begin{array}{ccccccccc}
\frac 45 & 0 & 0 & 0 & 0 & 0 & 0 & 0 & 0 \\
0 & -\frac 34 & 0 & 0 & 0 & 0 & 0 & 0 & 0 \\
0 & 0 & \frac 25 & 0 & 0 & 0 & 0 & 0 & 0 \\
0 & 0 & 0 & -\frac 14 & 0 & 0 & 0 & 0 & 0 \\
0 & 0 & 0 & 0 & 0 & 0 & 0 & 0 & 0 \\
0 & 0 & 0 & 0 & 0 & \frac 14 & 0 & 0 & 0 \\
0 & 0 & 0 & 0 & 0 & 0 & -\frac 25 & 0 & 0 \\
0 & 0 & 0 & 0 & 0 & 0 & 0 & \frac 34 & 0 \\
0 & 0 & 0 & 0 & 0 & 0 & 0 & 0 & -\frac 45
\end{array}
\right) $\\
\\
\,

$M_{9+}M_{9-}+M_{9-}M_{9+}=\left(
\begin{array}{ccccccccc}
\frac 45 & 0 & 0 & 0 & 0 & 0 & 0 & 0 & 0 \\
0 & \frac{17}{20} & 0 & 0 & 0 & 0 & 0 & 0 & 0 \\
0 & 0 & \frac 12 & 0 & 0 & 0 & 0 & 0 & 0 \\
0 & 0 & 0 & \frac{13}{20} & 0 & 0 & 0 & 0 & 0 \\
0 & 0 & 0 & 0 & \frac 25 & 0 & 0 & 0 & 0 \\
0 & 0 & 0 & 0 & 0 & \frac{13}{20} & 0 & 0 & 0 \\
0 & 0 & 0 & 0 & 0 & 0 & \frac 12 & 0 & 0 \\
0 & 0 & 0 & 0 & 0 & 0 & 0 & \frac{17}{20} & 0 \\
0 & 0 & 0 & 0 & 0 & 0 & 0 & 0 & \frac 45
\end{array}
\right) $\\
\\
\,

$Q_{9}=\left(
\begin{array}{ccccccccc}
0 & 0 & 0 & 0 & 0 & 0 & 0 & 0 & 0 \\
0 & \frac{15}{16} & 0 & 0 & 0 & 0 & 0 & 0 & 0 \\
0 & 0 & 7 & 0 & 0 & 0 & 0 & 0 & 0 \\
0 & 0 & 0 & \frac 43 & 0 & 0 & 0 & 0 & 0 \\
0 & 0 & 0 & 0 & 3 & 0 & 0 & 0 & 0 \\
0 & 0 & 0 & 0 & 0 & \frac 74 & 0 & 0 & 0 \\
0 & 0 & 0 & 0 & 0 & 0 & \frac 53 & 0 & 0 \\
0 & 0 & 0 & 0 & 0 & 0 & 0 & 0 & 0 \\
0 & 0 & 0 & 0 & 0 & 0 & 0 & 0 & 1
\end{array}
\right) \;\;\;\Longrightarrow
\;\;\;M_{9+}M_{9-}+Q_{9}M_{9-}M_{9+}=\frac 45{\Bbb I}_{9}$

\end{document}